\documentclass[preprint,showpacs,preprintnumbers,amsmath,amssymb]{revtex4}
\usepackage{graphicx}
\usepackage{dcolumn}
\usepackage{bm}
\begin{document}

\title{Statistical Self-Similar Properties of Complex Networks}
\author{Chang-Yong Lee}
\email{clee@kongju.ac.kr}
\affiliation{The Department of Industrial Information, Kongju National
  University, Chungnam, 340-702 South Korea}
\author{Sunghwan Jung}
\affiliation{Applied Mathematics Laboratory, Courant Institute for
  Mathematical Science, New York University, New York, NY 10012 USA}
\begin{abstract}
It has been shown that many complex networks shared distinctive
features, which differ in many ways from the random and the regular
networks. Although these features capture important characteristics of
complex networks, their applicability depends on the type of
networks. To unravel ubiquitous characteristics that complex networks
may have in common, we adopt the clustering coefficient as the
probability measure, and present a systematic analysis of various
types of complex networks from the perspective of statistical
self-similarity. We find that the probability distribution of the 
clustering coefficient is best characterized by the multifractal;
moreover, the support of the measure had a fractal dimension. These
two features enable us to describe complex networks in a unified way;
at the same time, offer unforeseen possibilities to comprehend complex
networks. 
\end{abstract}
\maketitle 

\section{Introduction}
The structure of complex systems across various disciplines can be  
abstracted and conceptualized as networks (or graphs) of nodes and links to
which many quantitative methods can be applied so as to extract any
characteristics embedded in the system~\cite{newmansiam}. 
Numerous complex systems have been expressed in
terms of networks, and have often been categorized by the research
field, such as social~\cite{amaral,newmansocial},
technological~\cite{fal,albertweb}, and biological
networks~\cite{guelzin,jeong}, to name a few. 

It was shown that many complex networks had distinctive
global features in common, including the small-world~\cite{watts} and
the scale-free~\cite{scale} properties. These uncovered
characteristics distinguish complex networks from the random and the regular
networks in that the average path between any two nodes is shorter
while maintaining highly clustered connections, and that the degree of
nodes approximately follows a power law distribution. 
In addition to the global characteristics, investigations on
complex networks at the local level have been directed to
reveal local characteristics from the
perspective of finding patterns of interconnections among
nodes. Notable examples include the network motif~\cite{milo}, the
(dis)assortativity~\cite{assort,maslov}, and the topological modules
in the metabolic~\cite{ravasz} as well as the protein
interaction networks~\cite{maslov}.  
Motivated by these characteristics, numerous models for the network
growth and evolution have been proposed to 
understand the underlying mechanism of complex networks. 

To gain further understanding of complex networks, we investigate
local features of the networks from the perspective of the statistical 
self-similarity. This may provide us with not only  
deeper insight into complex networks, but a unified way of 
describing them. To this end, we focus on the clustering
coefficient~\cite{watts} of a node $i$, defined as 
\begin{equation}
C_{i}= \frac{n_{i}}{k_{i}~(k_{i}-1)}~ ,
\end{equation}
where $n_{i}$, $0 \le n_{i} \le k_{i}(k_{i}-1)$,
is the number of directed links (an undirected link counts twice)
among $k_{i}$ nearest neighbor nodes of the node $i$. It is a measure of the 
interconnectivity among nearest neighbors of a node, or the
modularity~\cite{ravasz}, thus can be a quantity representing the local
structure of the network. 

The clustering coefficient has been analyzed from the perspective of
the degree correlation. It was found that the clustering coefficient
correlated with the degree in some complex networks. In particular,
there is a power law correlation between the clustering coefficient
and the degree for the deterministic scale-free network
model~\cite{doro}, the Internet~\cite{vazquez}, and metabolic
networks~\cite{ravasz}. It is a form of $\langle C \rangle(k) \sim
k^{- \delta}$, where $\langle \cdot \rangle$ represents the average
over the same degree, and $\delta$ depends on the type of
networks. However, a similar analysis of the clustering coefficient
reveals that the power law correlation is not manifest to other types
of networks, such as the protein interaction and social networks. As
shown in Fig.~1, the clustering coefficient of the film actor network
correlates with the degree not with a power-law, but exponentially (Fig.~1A);
while the neural network has an approximate linear correlation between the
clustering coefficient and the degree (Fig.~1B). Moreover, there is no
evident correlation in the protein networks (Fig.~1C, 1D). 

This finding suggests that in general the clustering coefficient is
not a simple quantity which can be related for some common ground to
other quantities, such as the degree. Thus, it is desirable to
analyze the clustering coefficient beyond the degree correlation.
In this paper, we focus on the systematic analysis of the clustering
coefficient in the wide classes of complex networks~\cite{data},
together with theoretical models for the complex networks, such as the
random~\cite{er}, the scale-free~\cite{scale}, and the small-world
networks~\cite{watts}. 
\section{Data Description}
We analyzed the clustering coefficient in the complex networks of the
film actor network, WWW, the scientific collaboration network,
metabolic networks, protein interaction networks, the neural network,
and the Internet of both Autonomous System (AS) and the router
levels; together with models for the random, the scale-free, and the
small-world networks. For directed networks such as metabolic networks
and WWW, we distinguish the network into the directionality (in-degree and
out-degree) and carry out separate analysis. The source of the network
data is in Ref.~\cite{data}, and more information
of each network is the following. \\
{\bf Film actor:}
  An actor represents a node, and actors casted in the same movie are
  linked. (374511 nodes and 2445818 undirected links) \\
{\bf Scientific collaboration:}
  Each author corresponds to a node and co-authors are linked among
  others in the Los Alamos E-print Archive between 1995 and 1999
  (inclusive) in the field of the condensed matter. (13861 nodes and 89238
  undirected links) \\  
{\bf Internet of Autonomous Systems (AS) level:}
  Each autonomous system represents a node, and a physical network connection
  between nodes represents a link. (6474 nodes and 25144 undirected links) \\
{\bf Internet of router level:}
  The Internet connection at the router level. The data is collected
  by the Mercator (\url{http://www.isi.edu/scan/mercator/}), a program
  that infers the router-level connectivity 
  of the Internet. (284772 nodes and 898456 undirected links) \\
{\bf WWW:}
  World Wide Web connection network for \url{http://www.nd.edu}. Each HTML
  document represents a node, connected directionally by a hyperlink
  pointing from one document to another. It is a directed 
  network of 325729 nodes and 1469679 directed links. \\
{\bf Metabolic networks:} 
  Metabolic networks of six organisms, two for
  each domain, are analyzed. They are {\it Archaeoglobus fulgidus} (459
  nodes and 2155 directed links) and {\it Methanobacterium
  thermoautotrophicum} (399 nodes and 1937 directed links) in Archae;
  {\it Escherichia coli} (698 nodes and 3747 directed links) and {\it
  Salmonella typhi} (735 nodes and 3882 directed links) in Bacteria;
  {\it Saccharomyces cerevisiae} (511 nodes and 2690  directed links)
  and {\it Caenorhabditis elegans} (413 nodes and 2061 directed links)
  in Eukaryote. Note that the metabolic network is a directed
  network. \\
{\bf Protein interaction networks:}
  We have analyzed protein interaction networks of six organisms. They
  are {\it Saccharomyces cerevisiae} (4687 nodes and
  30312 undirected links), {\it Escherichia coli} (145 nodes and 388
  undirected links), {\it Caenorhabditis elegans} (2386 nodes and 7650
  undirected links), {\it Drosophila melanogaster}  (6926 nodes and
  41490 undirected links), {\it Helicobacter pylori} (686 nodes and
  2702 undirected links), and {\it Homo sapiens} (563 nodes and 1740
  undirected links)  \\
{\bf Neural network:} 
  Somatic nervous system of {\it Nematode C. elegans} except that in
  the pharynz is considered. A link joins two nodes representing
  neurons, if they are connected by either a synapse or a gap
  junction. All links are treated as undirected. (265 nodes and 3664
  undirected links) \\
\section{Multifractality of complex networks}
The set of $C_{i}$ for
each network can be used to form a probability 
distribution (Fig.~2). As shown in Fig.~2A-2D, the distribution
of $C_{i}$ in complex networks differ considerably
from that of the random 
network (Fig.~2F). Probability distributions for complex networks
bring out high irregularity of various intensities in different
clustering coefficients, developing a long tail extending to either large
(Fig.~2A-2C) or small (Fig.~2D) values of the 
clustering coefficient. This suggests that not a few but many, possibly
infinite, parameters may be needed to characterize the distribution. 
To quantify the variation in the distribution, a continuous spectrum
of scaling indices has been proposed~\cite{halsey}.
For the spectrum that quantifies the inhomogeneity in the probability
distribution, we utilize the clustering coefficient as 
the probability measure, and analyze the distribution from the
perspective of the statistical self-similarity, the
multifractal~\cite{stanley,paladin}.  

The multifractal, which is not necessarily related to geometrical 
properties~\cite{benzi}, is a way to describe different self-similar
properties for different ``regions'' in an appropriate set (in our
case, different values of the clustering coefficient), and applied,
for instance, to the fully developed
turbulence~\cite{mandelbrot1,benzi}, which is one of the most common 
examples of complex systems. It consists of spectra displaying
the range of scaling indices and their density in the set, thus has
been used to explain richer and more complex scaling behavior of a
system than the case in the critical phenomena.  
The multifractal can be accomplished by examining the scaling 
properties of the measure characterized by the singularity strength 
$\alpha$ and its fractal dimension $f(\alpha)$, which
roughly indicates how often a particular value of $\alpha$
occurs~\cite{halsey}. In practice, $\alpha$ and $f(\alpha)$ are
customarily obtained from the Legendre transformation of $q$ and
$D_{q}$, via
\begin{equation}
\alpha=\frac{d}{dq} \{ (q-1) D_{q} \}~,
\end{equation}
and
\begin{equation}
f(\alpha)=q\alpha-(q-1)D_{q}~,
\end{equation}
where $D_{q}$ is the generalized correlation dimension often estimated
by the correlation integral method~\cite{hent}. It is the   
quantity for anomalous scaling law whose value depends on different
moment $q$. $D_{q}$ is defined as  
\begin{equation}
D_{q}=\lim_{R\rightarrow 0} \frac{1}{q-1} \frac{\ln S_{q}(R)}{\ln
  R}~,
\label{basic}
\end{equation}
where $S_{q}(R)$ is known as the correlation sum (or correlation
integral), and it is given, using Heaviside function $\Theta$, as 
\begin{equation}
S_{q}(R)=\frac{1}{M}\sum_{j=1}^{M}\left\{ \frac{1}{M-1} \sum_{k=1, k\ne
      j}^{M} \Theta\left(R-\left|C_{j}-C_{k}\right|\right)
      \right\}^{q-1} ,
\end{equation}
where $M$ is the number of nodes and $C_{i}$ is the clustering
coefficient of node $i$.  
The spectrum $D_{q}$ may be smoothed before transforming into $\alpha$
and $f(\alpha)$ to avoid the contradiction of the general
property of $D_{q}$, i.e., $D_{q} \le D_{q^{\prime}}~~\mbox{if}~~
q^{\prime} \le q$. 

There is a difficulty in taking the limit $R\rightarrow 0$ in
Eq.~(\ref{basic}) for a finite number of data points. Due to the
finiteness, there always exists the minimum distance of
$\left|C_{j}-C_{k}\right|$. Thus, when $R$ is less than the minimum
distance, the correlation sum becomes zero and no longer scales with
$R$. Therefore, in practice, the generalized dimension $D_{q}$ is determined by
plotting $\ln S_{q}(R)/(q-1)$ as a function of $\ln R$ and estimating
the slope within an appropriate scaling region using a least square
fit. The error associated with the fit can be obtained as a
statistical uncertainty based on fitting a straight line in the scaling region.

Fig.~3 displays the estimated $f(\alpha)$ versus $\alpha$ for various complex
networks. As shown in Fig.~3A-3D, the infinitely many different
fractal dimensions, manifested by the shapes of $f(\alpha)$, suggest
that the measure is a multifractal, and thus, cannot be explained by a
single fractal dimension. All of the complex networks we have
examined, except for the neural network of {\it Caenorhabditis
  elegans}, form multifractals irrespective of their global
characteristics, such as the number of nodes and their degree
distributions. Furthermore, for all complex networks we have analyzed,
the average and standard deviation of the most probable singularity
strength $\alpha_{0}$, where $f(\alpha)$ takes its maximum value, are
$\langle \alpha_{0} \rangle = 1.2 \pm 0.3$; those of $f(\alpha_{0})$
are $\langle f(\alpha_{0}) \rangle =0.8 \pm 0.1$.

The multifractal observed in complex networks implies that the
distribution of clustering coefficients can be described as
interwoven sets of singularities of strength $\alpha$, each of which
has its corresponding fractal dimension $f(\alpha)$~\cite{halsey}. In
our case, this implies that different values (or range of values) 
of the clustering coefficient may have different fractal dimensions. From
the viewpoint of network dynamics in which rewiring and/or adding
new nodes and links are involved, the multifractal suggests that as a
network grows, nodes of large clustering coefficients change their
clustering coefficients by a factor that differs from nodes of small
clustering coefficients change theirs. 

The different rate of changing the clustering
coefficient may stem from two sources (or modes): the degree of a node 
$k$ and the corresponding interconnectivity $n$. For a fixed $k$, the
clustering coefficient depends only on the interconnectivity $n$, so
that $C \sim n$. In this case, the dynamics (via rewiring and/or
adding new links) drive networks in such a way that different values
of $n$ are not equally probable; rather, some values of $n$ are more
probable than others. This assertion is further supported by the fact
that as $k$ increases, the number of distinct $n$ does not increase
quadratically in $k$, and that $n$ and $k$ are linearly correlated
(see below). For a fixed $n$, $C \sim k^{-2}$. Thus, the addition of
new links to higher degree nodes is more likely to drop their
clustering coefficient much faster than that of new links to lower degree
ones. Therefore, the dynamics of complex networks can be characterized
by an evolution via interplay between the two sources. 

Contrary to most complex networks, the irregularity of the
distribution is absent in the neural network of {\it Caenorhabditis
  elegans} (Fig.~2E). The estimate $D_{q}\approx 0.83 < 1$ is
independent of $q$, resulting in $\alpha=f(\alpha)\approx 0.83$. Thus,
the measure is not a multifractal, rather it can be characterized by a single
fractal dimension. The absence of the multifractality
is probably due to the biologically intrinsic property of the
neuron. The geometric character of the neuron imposes a
constraint on the number of synaptic contacts (i.e. links), leaving no
room for the irregularity of the distribution~\cite{amaral,neural}.  

Typically, $f(\alpha)$ satisfies $0 < f(\alpha) < D_{0}$, where $D_{0}$ is
the dimension of the support of the measure, which is the set of clustering
coefficients without their relative frequencies. We find
$D_{0} < 1$ for all complex networks (Fig.~3A-3D), indicating that 
supports of the measure have fractal dimensions, just like the Cantor 
set. This means that forbidden regions are embedded in the range of
allowed clustering coefficients so that some values are highly suppressed. 
A possible cause of the suppression might stem from the correlation between
the degree $k$ of a node and its corresponding interconnectivity $n$. 
A simple statistic, such as the
Pearson's correlation coefficient $r$, ranging $-1 < r < 1$, can be 
used to quantify the correlation, as it reflects to what extent the two
quantities are linearly correlated. The result (Fig.~4A) reveals that
complex networks have higher linear correlation than the
random network. Moreover, some complex networks, such as metabolic
networks and the Internet of AS level, disclose strong linear correlations
($r > 0.95$).  

For metabolic networks and the Internet of AS level in which the 
degree of a node and its interconnectivity among its nearest neighbor
nodes are strongly correlated, we ask how the next nearest neighbor
nodes are interconnected, and whether the distribution of the
corresponding clustering coefficients maintains the
multifractality. To this end, we extend the definition of the
clustering coefficient to the next nearest nodes by including the next
nearest nodes for both the degree of a node $k$ and its
interconnectivity $n$, which is the number of links between two next
nearest nodes. The irregularity and the long-tail 
characteristics are again found in the distribution of the extended
clustering coefficient, suggesting the existence of the
multifractal. As shown in Fig.~4B-4D, the probability distribution 
of the extended version of the measure can again be characterized by
the multifractal, indicating that the statistical self-similarity is
not necessarily restricted to the local interconnectivity.
By including the next nearest neighbor nodes to the definition of the 
clustering coefficient, more distinct values of the clustering
coefficient are possible than that of the nearest neighbor
nodes. This can be expected since the extended version of the
clustering coefficient of a node includes an average over the
clustering coefficients of its nearest neighbor nodes, partly
smoothing out the irregularity. This is also
manifested by the fact that $D_{0}^{\ast}$, the dimension of the
support of the extended clustering coefficient, is bigger than
corresponding $D_{0}$. 

Based on the multifractal found in complex networks, for  
comparison, we carried out similar analysis to models of the random,
the scale-free, and the small-world networks.
In the case of the random network, the clustering coefficients are
smoothly distributed, by having a ``bell-shape'' (Fig.~2F). Furthermore,
$D_{q}\approx 1.0$ for all $q$, indicating that there is no
self-similarity. This can be expected since the support of the
measure can be regarded as a line, which is one dimensional.  

The distribution of the clustering coefficient for the scale-free network
shows the irregularity, similar to the case of complex networks;
furthermore the 
distribution can be described by the multifractal (Fig.~3E). From
simulation results with various different parameters for the network,
however, we found that not only the most probable singularity
strength $\alpha_{0}$ ($0.64 < \alpha_{0} < 0.76$), 
but the dimension of the support $D_{0}$ ($0.45 < D_{0} < 0.50$) 
is smaller than that of complex networks, suggesting that more
severe restriction is imposed on the possible values of the clustering
coefficient. According to the model, nodes
of higher degree are preferred to have additional links rather than those
of lower degree. This preferential attachment leaves the clustering
coefficient of high degree nodes to decrease much faster than that of
low degree 
ones. Thus when the number of nodes is doubled, for instance, the 
clustering coefficient of high degree nodes changes by a factor
different from that of low degree nodes, analogous to the kinetics of
the diffusion-limited aggregation~\cite{meakin}. 

For the small-world network, the rewiring probability $p$ dictates
both the irregularity in the distribution and the multifractality. For a small
rewiring probability (say, $p=0.01$), the multifractal emerges
(Fig.~3F); however the dimension of the support is $D_{0}\approx 1.0$,
implying that the set of the measure entirely covers the space of the
clustering coefficient. As the rewiring probability increases, the range of
the clustering coefficients becomes smaller and the degree of inhomogeneity 
decreases, and then the network finally becomes a random network as we can
easily anticipate.  
\section{Summary and Conclusion} 
Based on the irregularity of intensities in the probability
distribution of the clustering coefficient, 
we regarded the clustering coefficient as the probability measure and
analyzed the clustering coefficient of various types of complex
networks from the perspective of the statistical self-similarity. We
found that the probability measure and the support of the measure can be
characterized by the multifractal and the fractal,
respectively. Furthermore, for complex networks having strong linear
correlation between the degree and the interconnectivity, the
multifractality extends into the clustering coefficient of the next
nearest neighbor nodes. These characteristics are unique to the real
complex networks and cannot be found in the
random network. From the aspect of the multifractality, models
of the scale-free and the small-world differ from real networks in the
distribution of the singularity strength $f(\alpha)$.

The statistical self-similarity in the distribution of the clustering
coefficient can be served as a general characteristic of complex
networks; at the same time, giving a further insight into the
understanding of complex networks. The multifractality shared by
different complex networks suggests that a similar law may govern 
the diverse complex networks of nature as well as artificiality.
Furthermore, it can be used to classify the complex networks, and
serves as a ``touchstone'' of proposed models for complex networks.  
\begin{acknowledgments} 
We like to thank M. Newman for providing us with the scientific
collaboration data. We also appreciate the open source of various
complex network data available at
\url{http://www.nd.edu/~networks/}. This work was supported by the
Korea Research Foundation Grant funded by the Korean Government (MOEHRD)
(KRF-2005-041-H00052).
\end{acknowledgments}

\newpage
\begin{figure}
\centerline{
\includegraphics{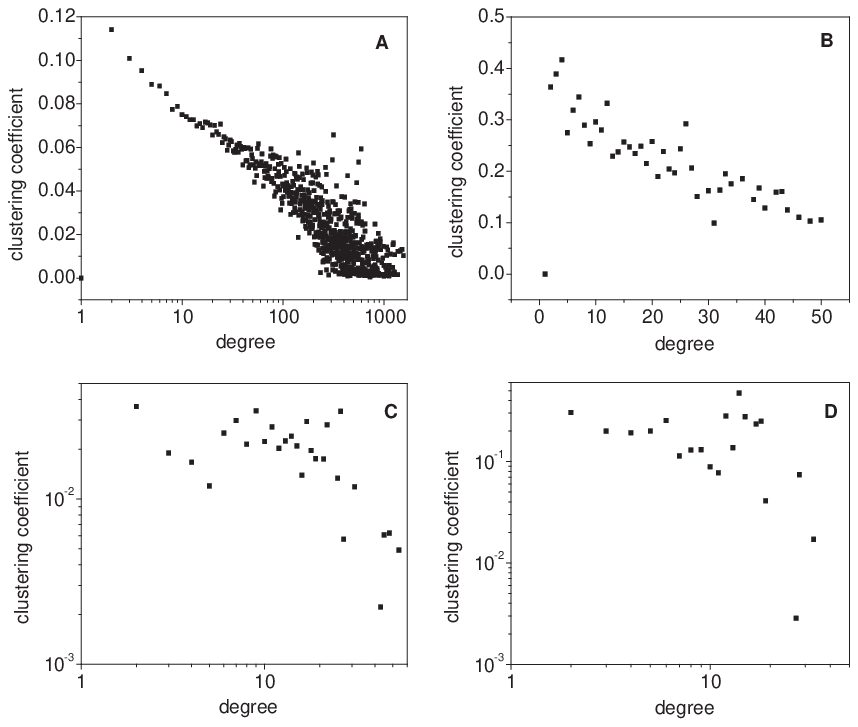}
}
\caption{Plots of the clustering coefficient averaged over all nodes
  of the same degree versus the degree for selected complex
  networks. (A) the film actor network, (B) the neural network of {\it
  Caenorhabditis elegans}, (C) the protein interaction network of {\it
  Helicobacter pylori}, (D) the protein interaction network of {\it
  Homo sapiens}. Note that the abscissa of (A), and the abscissa as
  well as the ordinate of (C) and (D) are in log scale.}
\end{figure}
\begin{figure}
\centerline{
\includegraphics{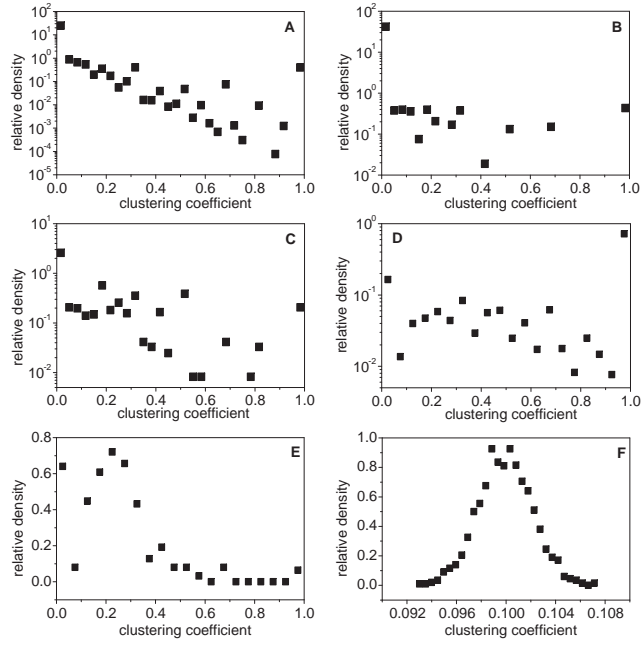}
}
\caption{Probability distributions of the clustering coefficient for
  selected complex networks as examples. (A) the film actor network, (B)
  the protein interaction network for {\it Caenorhabditis elegans}, (C) the
  metabolic network (in-degree) of {\it Escherichia coli}, (D) the
  scientific collaboration network, (E) the neural network, and (F)
  the random network of 2000 nodes with the connection probability
  $p=0.1$. Note that ordinates of (A)-(D) are in log scale for 
  the display purpose.}
\end{figure}
\begin{figure}
\centerline{
\includegraphics{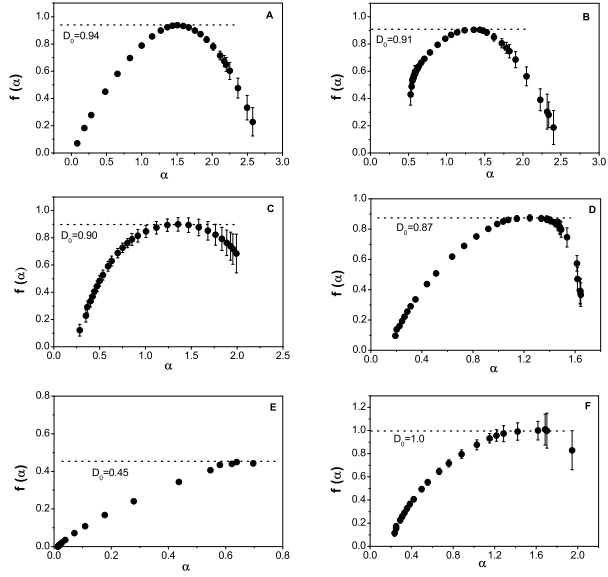}
}
\caption{$f(\alpha)$ versus $\alpha$ for selected complex networks as
  examples. Error-bars are the root mean square in estimating
  $f(\alpha)$, and $D_{0}$ (the maximum of $f(\alpha)$) is the 
  dimension of the support of the measure. (A) WWW (in-degree), (B) 
  the scientific collaboration network (cond-mat), (C) the
  metabolic network (out-degree) of {\it Escherichia coli}, (D) the
  protein interaction network of {\it Saccharomyces
  cerevisiae}, (E) the model of the scale-free network with 10000 nodes,
  (F) the small-world model of the 
  rewiring probability $p=0.01$.}
\end{figure}
\begin{figure}
\centerline{
\includegraphics{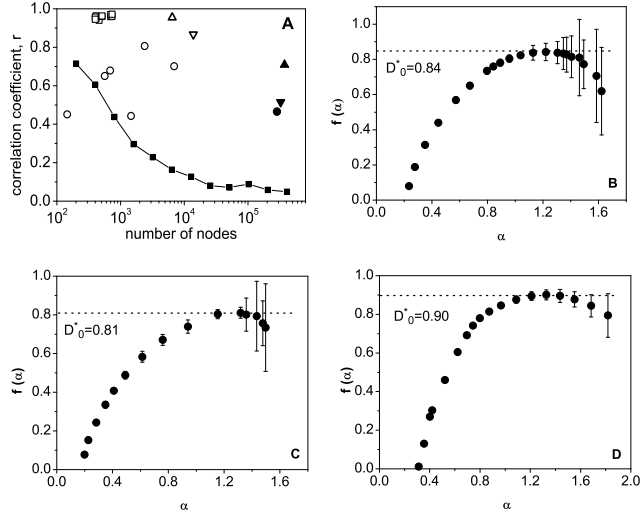}
}
\caption{(A): The linear correlation coefficients $r$ of complex networks
  plotted against the number of nodes are compared with the
  random networks of different numbers of nodes: the
  metabolic networks ($\Box$), the Internet of AS level ($\triangle$),
  the scientific collaboration network ($\triangledown$), protein
  interaction networks ($\bigcirc$), the actor network
  ($\blacktriangle$), WWW ($\blacktriangledown$), the Internet of
  router level ($\bullet$), and the random network
  ($\blacksquare$). For random networks, 
  each node has on average 10 links. (B)-(D): $f(\alpha)$ versus
  $\alpha$ for selected organisms in the metabolic networks and the Internet of
  the AS level. $D^{\ast}_{0}$ represents the dimension of the support
  for the extended clustering coefficient. (B) the 
  metabolic network of {\it Archaeoglobus fulgidus} (in-degree), (C) the
  metabolic network of {\it Caenorhabditis elegans} (in-degree), (D)
  the Internet of AS level. Note that $D^{\ast}_{0}$ for (B)-(D) are larger
  than corresponding $D_{0}$ for the case of the nearest neighbor,
  where, from 3B to 3D, $D_{0}=0.79, 0.73, 0.87$, respectively.}
\end{figure}
\end{document}